\documentclass[twocolumn,times]{aastex631}

\let\tablenum\relax

\usepackage[T1]{fontenc}
\usepackage{amsmath}
\usepackage{amssymb}
\usepackage{bm}
\usepackage{lipsum}
\usepackage{siunitx}

\newcommand{\munull}{\mu_{0}}
\newcommand{\vAlfven}{v_{\mathrm{A}}}

\newcommand{\Eplusrect}{E_{+}^{*}}

\newcommand{\Eminusrect}{E_{-}^{*}}
\newcommand{\Epm}{E_{\pm}}
\newcommand{\Epmrect}{E_{\pm}^{*}}

\newcommand{\rE}{r_{\mathrm{E}}}
\newcommand{\rErect}{r_{\mathrm{E}}^{*}}
\newcommand{\crosshel}{\sigma_{\mathrm{c}}}
\newcommand{\crosshelrect}{\sigma_{\mathrm{c}}^{*}}

\newcommand{\resenerg}{\sigma_{\mathrm{r}}}
\newcommand{\rA}{r_{\mathrm{A}}}

\newcommand{\KHI}{|\Delta v|/v_{\mathrm{A}}}
\newcommand{\Bcomp}{\delta|B|/\delta B}

\received{October 29, 2022}
\revised{February 17, 2023}
\accepted{March 2, 2023}

\shortauthors{Soljento et al.}

\graphicspath{{./}{img/}}

\begin{document}

\title{%
    Imbalanced Turbulence Modified by Large-scale Velocity Shears in the Solar
    Wind}

%% you desire. In v6.31 you can include a footnote in the title.

%% A significant change from earlier AASTEX versions is in the structure for 
%% calling author and affiliations. The change was necessary to implement 
%% auto-indexing of affiliations which prior was a manual process that could 
%% easily be tedious in large author manuscripts.
%%
%% The \author command is the same as before except it now takes an optional
%% argument which is the 16 digit ORCID. The syntax is:
%% \author[xxxx-xxxx-xxxx-xxxx]{Author Name}
%%
%% This will hyperlink the author name to the author's ORCID page. Note that
%% during compilation, LaTeX will do some limited checking of the format of
%% the ID to make sure it is valid. If the "orcid-ID.png" image file is 
%% present or in the LaTeX pathway, the OrcID icon will appear next to
%% the authors name.
%%
%% Use \email to set provide email addresses. Each \email will appear on its
%% own line so you can put multiple email address in one \email call. A new
%% \correspondingauthor command is available in V6.31 to identify the
%% corresponding author of the manuscript. It is the author's responsibility
%% to make sure this name is also in the author list.

\correspondingauthor{Juska E. Soljento}
\email{juska.soljento@helsinki.fi}

\author[0000-0003-2495-8881]{Juska E. Soljento}
\affiliation{Department of Physics, University of Helsinki, Helsinki, Finland}

\author[0000-0002-4921-4208]{Simon W. Good}
\affiliation{Department of Physics, University of Helsinki, Helsinki, Finland}

\author[0000-0003-2555-5953]{Adnane Osmane}
\affiliation{Department of Physics, University of Helsinki, Helsinki, Finland}

\author[0000-0002-4489-8073]{Emilia K. J. Kilpua}
\affiliation{Department of Physics, University of Helsinki, Helsinki, Finland}

\begin{abstract}
    We have investigated how the degree of imbalance in solar wind turbulence is
        modified by large-scale velocity shears in the solar wind plasma.
    The balance between counterpropagating Alfvénic fluctuations, which interact
        nonlinearly to generate the turbulence, has been quantified by the cross
        helicity and Elsasser ratio.
    Velocity shears at a 30-min timescale were identified, with the shear
        amplitude defined in terms of the linear Kelvin--Helmholtz (KH)
        instability threshold.
    The shears were associated with 74 interplanetary coronal mass ejection
        (ICME) sheaths observed by the \emph{Wind} spacecraft at
        \SI{1}{\astronomicalunit} between 1997 and 2018.
    Typically weaker shears upstream of the sheaths and downstream in the ICME
        ejecta were also analyzed.
    In shears below the KH threshold, imbalance was approximately invariant or
        weakly rising with shear amplitude.
    Above the KH threshold, fluctuations tended toward a balanced state with
        increasing shear amplitude.
    Magnetic compressibility was also found to increase above the KH threshold.
    These findings are consistent with velocity shears being local sources of
        sunward fluctuations that act to reduce net imbalances in the
        antisunward direction, and suggest that the KH instability plays a role
        in this process.
\end{abstract}


%%%%%%%%%%%%%%%%%%%%%%%%%%%%%%%%%%%%%%%%%%%%%%%%%%%%%%%%%%%%%%%%%%%%%%%%%%%%%%%%
\section{Introduction}\label{sec:intro}

Properties of fluctuations at magnetohydrodynamic (MHD) scales in the solar
    wind are consistent with the presence of a turbulent cascade
    \citep[e.g.,][]{bruno2016,chen2016,kiyani2015}.
Solar wind turbulence is primarily Alfvénic in nature, and typically displays a
    predominance of antisunward-directed fluctuations
    \citep{belcher1971b,goldstein1999b}.
This antisunward imbalance decreases with radial distance
    \citep[e.g.,][]{roberts1987a,roberts1987b,marsch1990,bavassano2000} and also
    varies with solar wind type \citep{borovsky2012b,borovsky2019}.
At low frequencies, in the energy-containing \(f^{-1}\) range, the Alfvénic
    fluctuations are largely noninteracting, while at higher frequencies, in
    the inertial range, the counterpropagating fluctuations interact nonlinearly
    and drive a turbulent cascade \citep[e.g.,][]{bruno2016,chen2016}.

The minority sunward component of the fluctuations is thought to be generated
    locally in interplanetary space
    \citep[e.g.,][]{goldstein1995,bavassano1996,petrosyan2010}. 
Possible generation mechanisms include the parametric decay instability
    \citep[e.g.,][]{bowen2018,reville2018,malara2022,sishtla2022}, reflection
    off the radial gradient in the Alfvén speed \citep[e.g.,][]{chandran2011},
    and velocity shear \citep[e.g.,][]{roberts1992,goldstein1999a,stawarz2011}.
It should be noted that effects such as velocity shear and reflection from
    gradients are well-known features of the non-WKB transport of Alfvén waves
    that occurs in the solar wind \citep{heinemann1980,zhou1989,zhou1990}, and
    they have been shown to contribute to the driving of turbulence in both
    simulations \citep[e.g.,][]{zank1996,breech2008} and in situ studies
    \citep[e.g.,][]{roberts1987a,roberts1987b,bavassano1998}.
Multiple effects can be present at the same time and contribute to the
    generation of sunward fluctuations.
Out of the effects mentioned above, we focus here on the role of large-scale
    velocity shear.

Driving effects such as velocity shear are necessary, because as suggested by
    \citet{dobrowolny1980}, the imbalance between the counterpropagating
    Alfvénic fluctuations tends to increase in their absence.
This was shown to be the case in MHD simulations by \citet{pouquet1996}.
This, however, is in contrast with the observed decrease of imbalance with
    distance from the Sun.
Velocity shear was advanced as a possible candidate to explain the observed
    behavior of the fluctuations \citep{roberts1992}.
Subsequent studies by \citet{matthaeus2004} and \citet{breech2005} showed that
    introducing shear as a driver of turbulence could explain the observed
    radial evolution of the balance between fluctuations in the ecliptic as well
    as at higher latitudes.
The connection between velocity shear and the local generation of fluctuations
    is still an open question, with some finding a link between the two
    \citep[e.g.,][]{smith2011} and others not \citep[e.g.,][]{borovsky2010}.

Just as with hydrodynamic fluids \citep[e.g.,][]{rogers1992}, a shear
    interface in an MHD fluid can develop a Kelvin--Helmholtz (KH) instability,
    which can evolve into a vortex rollup.
In the solar wind, a magnetized KH instability is suppressed when the magnetic
    field parallel to the bulk motion of the plasma acts as a resisting force
    against the formation of vortices that would occur at the boundary of the
    shear layer \citep[for a recent review, see][]{faganello2017}.
If, however, the difference in speed across the interface is greater than the
    local Alfvén speed, i.e., if \(|\Delta v| > \vAlfven\), where \(\vAlfven =
    B/\sqrt{\munull\rho}\) and \(\rho\) is the ion mass density, then the
    magnetic field can no longer suppress the instability and vortex rollup may
    occur \citep{chandrasekhar1981, ruffolo2020}.  
    
 In this Letter, we directly examine for the first time the relationship
    between imbalance in the solar wind turbulence and large-scale shear
    amplitudes in terms of the KH instability, via a statistical study of 74
    sheath regions driven by interplanetary coronal mass ejections
    \citep[ICMEs;][]{kilpua2017,luhmann2020} observed by the \emph{Wind}
    spacecraft at \SI{1}{\astronomicalunit}.
The regions upstream and downstream of each sheath have also been examined.
Besides ICME sheaths being of general interest for studies of solar wind
    turbulence, the prevalence of strong shears that we have found in sheaths
    makes them a particularly useful environment for the present study.

This Letter is organized as follows:
    In Section~\ref{sec:observations}, the spacecraft data are described and
    key parameters are defined. In Section~\ref{sec:analysis}, the analysis is
    presented, with Section~\ref{ssec:example_event} presenting an example
    event that highlights some of the key parameters that are later studied
    in more detail.
Sections~\ref{ssec:parameter_distributions}
    and~\ref{ssec:velocity_shear_dependence} present the statistical analysis
    of the full 74 sheath set, and finally in
    Section~\ref{sec:discussion_conclusion} the results of the analysis are 
    discussed in detail and conclusions are presented.

%%%%%%%%%%%%%%%%%%%%%%%%%%%%%%%%%%%%%%%%%%%%%%%%%%%%%%%%%%%%%%%%%%%%%%%%%%%%%%%%
\section{Data and Methods}\label{sec:observations}

Magnetic field data from MFI~\citep{lepping1995}, and ion moments from
    the 3DP/PESA-L~\citep{lin1995}, both on board the \textit{Wind} spacecraft,
    were analyzed.
The event list consists of 74 sheath regions identified and studied earlier by
    \citet{kilpua2021a}.
The sheaths occurred between 1997 January and 2018 March, which approximately
    covers solar cycles 23 and 24.
The data were resampled to the same resolution (on average about
    \SI{3.1}{\second}), and any small data gaps were closed using linear
    interpolation.

Using measurements of the magnetic field, \(\bm{B}\), proton velocity,
    \(\bm{v}\), and proton number density, \(n_{\mathrm{p}}\), a number of
    parameters were calculated.
These include the linear KH instability parameter, \(\KHI\), used as a measure
    of velocity shear in the solar wind plasma.
In the numerator of \(\KHI\),%\vspace{-2ex}
    \begin{align}\label{eqn:velocity_shear}
        |\Delta v|
            &= |\bm{v}_{\perp}(t + \tau_{\mathrm{s}}) - \bm{v}_{\perp}(t)|
                \nonumber\\
            &= \sqrt{[v_{Y}(t + \tau_{\mathrm{s}}) - v_{Y}
                (t)]^{2} + [v_{Z}(t + \tau_{\mathrm{s}}) -
                v_{Z}(t)]^{2}},
    \end{align}
    where \(\tau_{\mathrm{s}}\) is the timescale over which velocity shear is
    taken to occur.
Only changes in the components perpendicular to the radial direction, i.e., GSE
    \(X\), are taken to contribute to velocity shear.
This is done to distinguish velocity shear from any radial compression of the
    plasma.
Here we investigate the possible link between large-scale velocity shear and
    turbulence at smaller scales, with \(\tau_{\mathrm{s}}\) chosen to be
    \SI{1800}{\second} (\SI{30}{\minute}).

The \citet{elsasser1950} variables, \(\bm{z}^{\pm}\), were also calculated.
They are defined as \(\bm{z}^{\pm} = \bm{v} \pm \bm{b}\), where \(\bm{b} =
    \bm{B}/ \sqrt{\munull\rho}\) is the magnetic field in Alfvén units.
The solar wind was taken to consist entirely of protons, such that \(\rho =
    m_{\mathrm{p}}n_{\mathrm{p}}\), where \(m_{\mathrm{p}}\) is the mass of a
    proton.
Fluctuations in the Elsasser variables correspond to Alfvénic wave packets
    propagating parallel (i.e., \(\bm{z}^{-}\) fluctuations) or antiparallel
    (i.e., \(\bm{z}^{+}\) fluctuations) to the background magnetic field
    \citep[e.g.,][]{goldstein1999b}.

The power spectral densities (PSDs) of \(\bm{v}\), \(\bm{b}\), and
    \(\bm{z}^{\pm}\), denoted \(E_{v}\), \(E_{b}\), and  \(E_{\pm}\),
    respectively, were determined using wavelet analysis~\citep{torrence1998}.
Morlet wavelets were used in the wavelet transforms.
Since \(\bm{v}\), \(\bm{b}\), and \(\bm{z}^{\pm}\) are vectors, wavelet
    transforms and PSDs of all three components of the vectors were calculated
    and summed together to obtain the PSDs of the full vector quantities.
For example, \(E_{v}\) is given by
    \begin{align}\label{eqn:power_spectral_density}
        E_{v}
            &= E_{v_{x}} + E_{v_{y}} + E_{v_{z}} \nonumber\\
            &= |\mathcal{W}_{v_{x}}|^{2} + |\mathcal{W}_{v_{y}}|^{2} +
                |\mathcal{W}_{v_{z}}|^{2},
    \end{align}
    where \(\mathcal{W}_{v_{i}}\) are the wavelet transforms of the velocity
    components, and \(|\mathcal{W}_{v_{i}}|^{2}\) give their PSDs.
The wavelet transforms were performed over the frequency range \num{e-3} to 
    \SI{e-2}{\hertz}, equivalent to wave periods ranging from \num{1.67} to
    \SI{16.7}{\minute}.
This frequency range is in the inertial range of MHD turbulence at
    \SI{1}{\astronomicalunit} \citep[e.g.,][]{kiyani2015}.

Using the PSDs, four turbulence parameters were calculated: the normalized
    cross helicity, \(\sigma_{\mathrm{c}}\), the Elsasser ratio,
    \(r_{\mathrm{E}}\), the normalized residual energy, \(\sigma_{\mathrm{r}}\),
    and the Alfvén ratio, \(r_{\mathrm{A}}\).
They are defined as \citep[e.g.,][]{bavassano1998,chen2013}
    \begin{alignat}{2}
        \crosshel
            &= \frac{E_{+} - E_{-}}{E_{+} + E_{-}},\quad
        &\rE
            &= \frac{E_{+}}{E_{-}}, \label{eqn:sigmac_rE}\\
        \resenerg
            &= \frac{E_{v} - E_{b}}{E_{v} + E_{b}},
        &\rA
            &= \frac{E_{v}}{E_{b}}. \label{eqn:sigmar_rA}
    \end{alignat}
Cross helicity and the Elsasser ratio quantify the balance of power between
    parallel and antiparallel fluctuations, while residual energy and the
    Alfvén ratio quantify the balance of power between velocity and magnetic
    field fluctuations.
In the following, cross helicity and residual energy always refer to the
    normalized quantities.

When calculating the statistical distributions of \(\crosshel\) and \(\rE\), the
    direction of the magnetic field was rectified \citep{bruno1985,roberts1987b}
    such that \(\bm{z}^{+}\) always corresponds to Alfvénic fluctuations
    propagating in the antisunward direction, and \(\bm{z}^{-}\) to Alfvénic
    fluctuations propagating in the sunward direction.
In rectification the magnetic field sign is flipped when its direction is
    outward from the Sun.
The boundaries between the outward and inward sectors are defined in relation to
    the nominal Parker spiral angle, \(\phi_{\mathrm{PS}}\), which was assumed
    to be \SI{44}{\degree} at \(\mathrm{L}_{1}\), where \emph{Wind} is located.
The rectified quantities are denoted by asterisks, e.g., \(\crosshelrect\)
    refers to rectified cross helicity.
In practice rectification means that in the outward sector \(\Epmrect = 
    E_{\mp}\), which implies that \(\crosshelrect = -\crosshel\) and
    \(\rErect =  \rE^{-1}\).
    
Also considered is the magnetic compressibility, \(\Bcomp\)
    \citep[e.g.,][]{chen2015,good2020a}, defined as
    \begin{equation}
        \frac{\delta|B|}{\delta B}
            = \frac{\delta|B|}{|\delta\bm{B}|}
            = \frac{||\bm{B}(t + \tau_{\mathrm{c}})| - |\bm{B}(t)||}{|\bm{B}(t +
                \tau_{\mathrm{c}}) - \bm{B}(t)|},
    \end{equation}
    where \(\tau_{\mathrm{c}}\) is the fluctuation timescale.
Compressibility quantifies the degree to which magnetic fluctuations involve
    compression, i.e., a change in the magnitude of the magnetic field rather
    than a rotation of the magnetic field vector.
Here \(\tau_{\mathrm{c}} = \SI{300}{\second}\) (\SI{5}{\minute}), which is near
    the middle of the studied fluctuations' period scale.
This allows for comparison of compressibility with other turbulence properties
    at a similar scale.

%%%%%%%%%%%%%%%%%%%%%%%%%%%%%%%%%%%%%%%%%%%%%%%%%%%%%%%%%%%%%%%%%%%%%%%%%%%%%%%%
\section{Analysis}\label{sec:analysis}

\subsection{An Example Event}\label{ssec:example_event}

\begin{figure*}
    \centering
    \includegraphics[width=\textwidth]{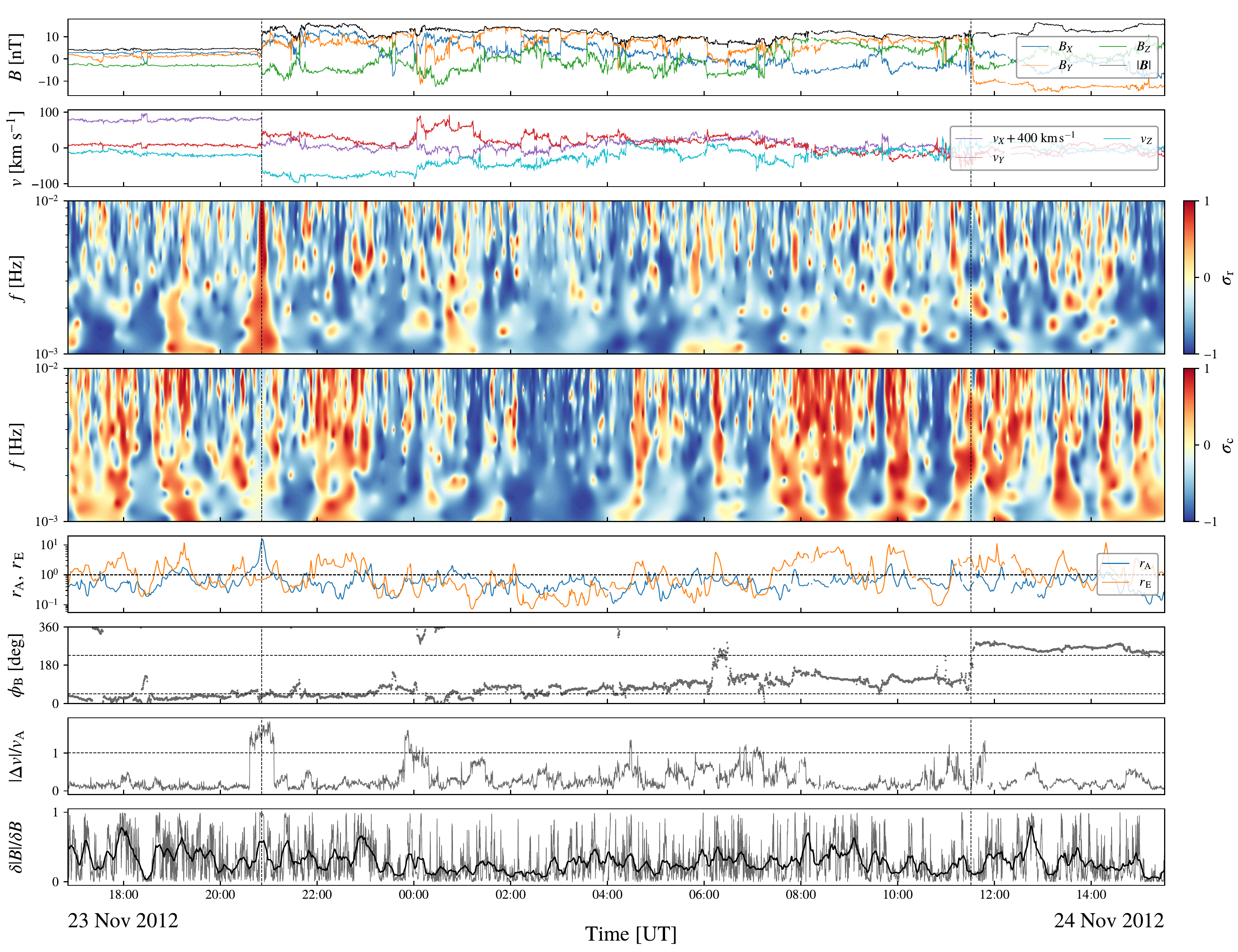}
    \caption{%
        An example ICME.
        From top to bottom, the panels show: the magnetic field components in
            GSE coordinates and the magnetic field magnitude; proton velocity
            components in GSE coordinates; normalized residual energy;
            normalized cross helicity; the Alfvén ratio and the Elsasser ratio;
            the magnetic field longitude angle in GSE coordinates; the
            magnitude of the difference in the nonradial velocity components
            normalized by the local Alfvén speed; and magnetic compressibility
            along with a 10-min running average (black curve) to show its
            overall evolution.
        In the longitude panel the horizontal dashed lines denote the boundaries
            between the sunward and antisunward sectors of the interplanetary
            magnetic field.
        The dashed vertical lines across all panels denote the shock and the
            ICME leading edge, respectively.}
    \label{fig:example_icme}
\end{figure*}
    
Figure~\ref{fig:example_icme} shows an example ICME observed by \emph{Wind} in
    2012 November.
The sheath is between the shock and the ejecta leading edge, which are indicated
    by dashed vertical lines.
The top two panels show the magnetic field and velocity components,
    respectively, and the third and fourth panels show the wavelet power
    spectra of residual energy, \(\resenerg\), and cross helicity,
    \(\crosshel\), respectively.

The third panel shows that in all three intervals (upstream wind, sheath, and
    ICME ejecta), \(\resenerg\) is overall negative, with low \(|\resenerg|\).
This is reflected in the fifth panel, which shows that the Alfvén ratio \(\rA\),
    averaged across frequency, is mostly below one.
Negative \(\resenerg\) (or \(\rA < 1\)) indicate a dominance of magnetic
    fluctuations over velocity fluctuations, and low \(|\resenerg|\) (or \(\rA
    \sim 1\)) indicates that the flow is highly Alfvénic \citep{bavassano1998}.
There is a clear positive peak in \(\resenerg\) spanning the entire frequency
    range at the shock.
A similar feature is evident in the residual energy spectra of the majority of
    the sheaths studied.
In the fifth panel \(\rA\) also peaks sharply at the shock.

The third panel from the bottom shows the magnetic field longitude,
    \(\phi_{\mathrm{B}}\).
The outward/inward sector boundaries are denoted by the horizontal dashed lines,
    the outward sector being between the lines and the inward sector outside
    them.

The cross helicity, \(\crosshel\), shown in the fourth panel indicates that in
    the upstream solar wind and the early part of the sheath there is no clear
    tendency for it being dominantly positive or negative. Note
    that here \(\crosshel\) and \(\rE\) have not been rectified to avoid
    discontinuities in the plots.
Before 11/24 00:00 UT the magnetic field is mostly in the inward sector, as
    seen in the longitude panel, which means that positive \(\crosshel\)
    corresponds to antisunward fluctuations dominating.
This flips when the magnetic field drifts from the inward to the outward sector
    sometime after 23:00 UT and stays there for the rest of the duration of the
    sheath.
Most of the middle part of the sheath has overall negative \(\crosshel\) (and
    \(\rE < 1\) as seen in the fifth panel), which now implies more
    antisunward fluctuations.
There is a patch of sunward fluctuations at the end of the sheath, from
    before 08:00 UT until after 10:00 UT.
In the ejecta the magnetic field again points toward the Sun and the
    \(\crosshel\) profile looks similar to the upstream profile.

\begin{figure}
    \centering
    \includegraphics[width=0.47\textwidth]{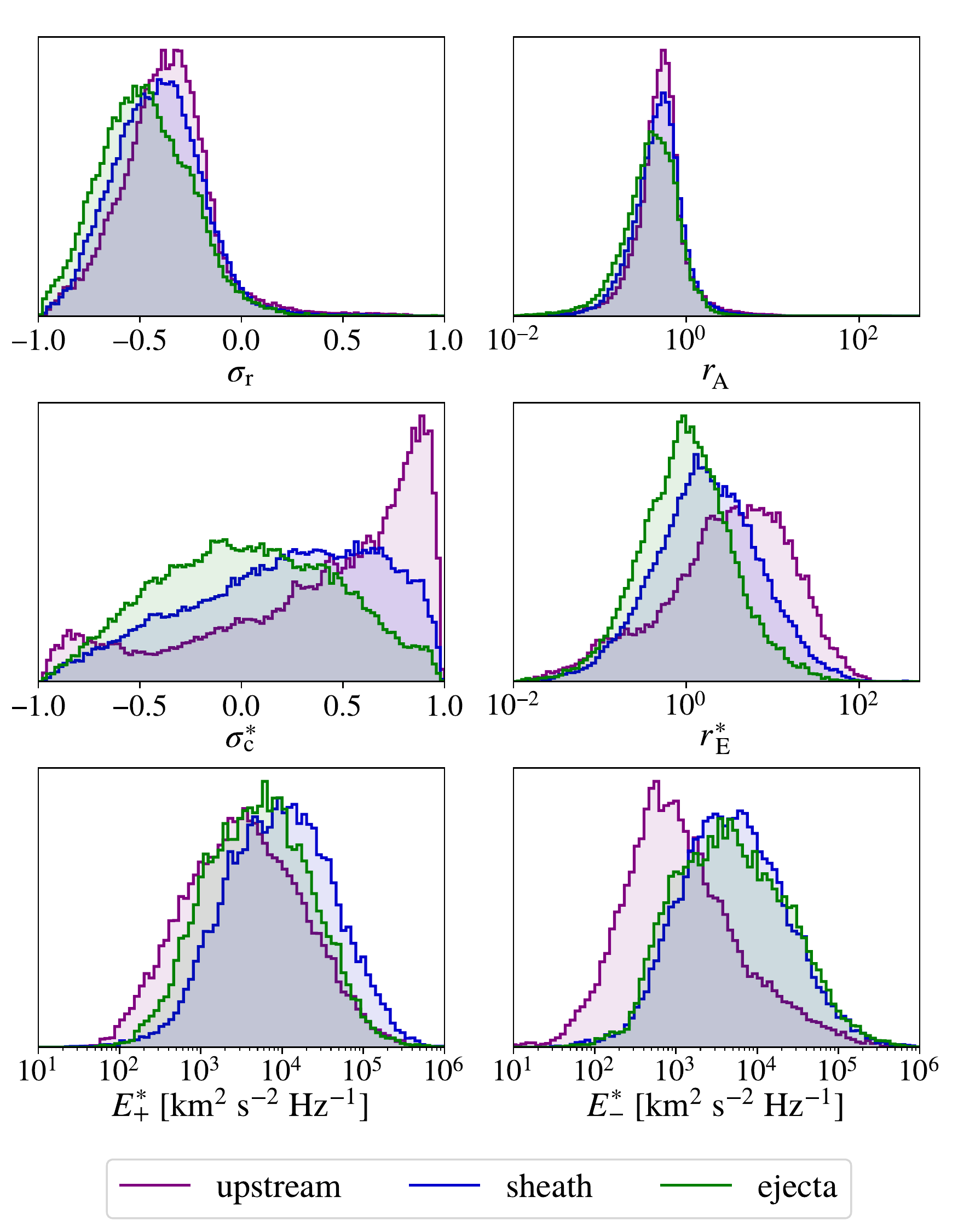}
    \caption{%
        Distributions of key turbulence parameters upstream of the shock
            (purple)‚ inside the sheath (blue), and in the ICME ejecta (green),
            summed across all 74 events.
        The residual energy, \(\resenerg\), and Alfvén ratio, \(\rA\), are on
            the top row, the rectified cross helicity, \(\crosshelrect\), and
            Elsasser ratio, \(\rErect\), in the middle row, and the rectified
            PSDs, \(\Epmrect\), of \(\bm{z}^{\pm}\) are on the bottom row.}
    \label{fig:1d_distributions}
\end{figure}

The second panel from the bottom shows the velocity shear amplitude in terms of
    the linear KH instability parameter, \(\KHI\).
Here and in the subsequent analysis, \(\Delta v\) and \(\vAlfven\) are
    calculated at the base data resolution (\(\sim\SI{3.1}{\second}\)).
Across the time interval shown, it can be seen that most shear amplitudes
    approaching or exceeding the KH threshold were located in the sheath. 

The bottom panel of Figure~\ref{fig:example_icme} shows the magnetic
    compressibility, \(\Bcomp\).
It fluctuated rapidly between high and low values, so a 10-min running average
    was calculated (black curve) to show its overall variation.
This average line also shows significant variability, but without any strong
    systematic differences between the sheath interval and upstream or
    downstream.

\begin{figure*}
    \centering
    \includegraphics[width=0.90\textwidth]{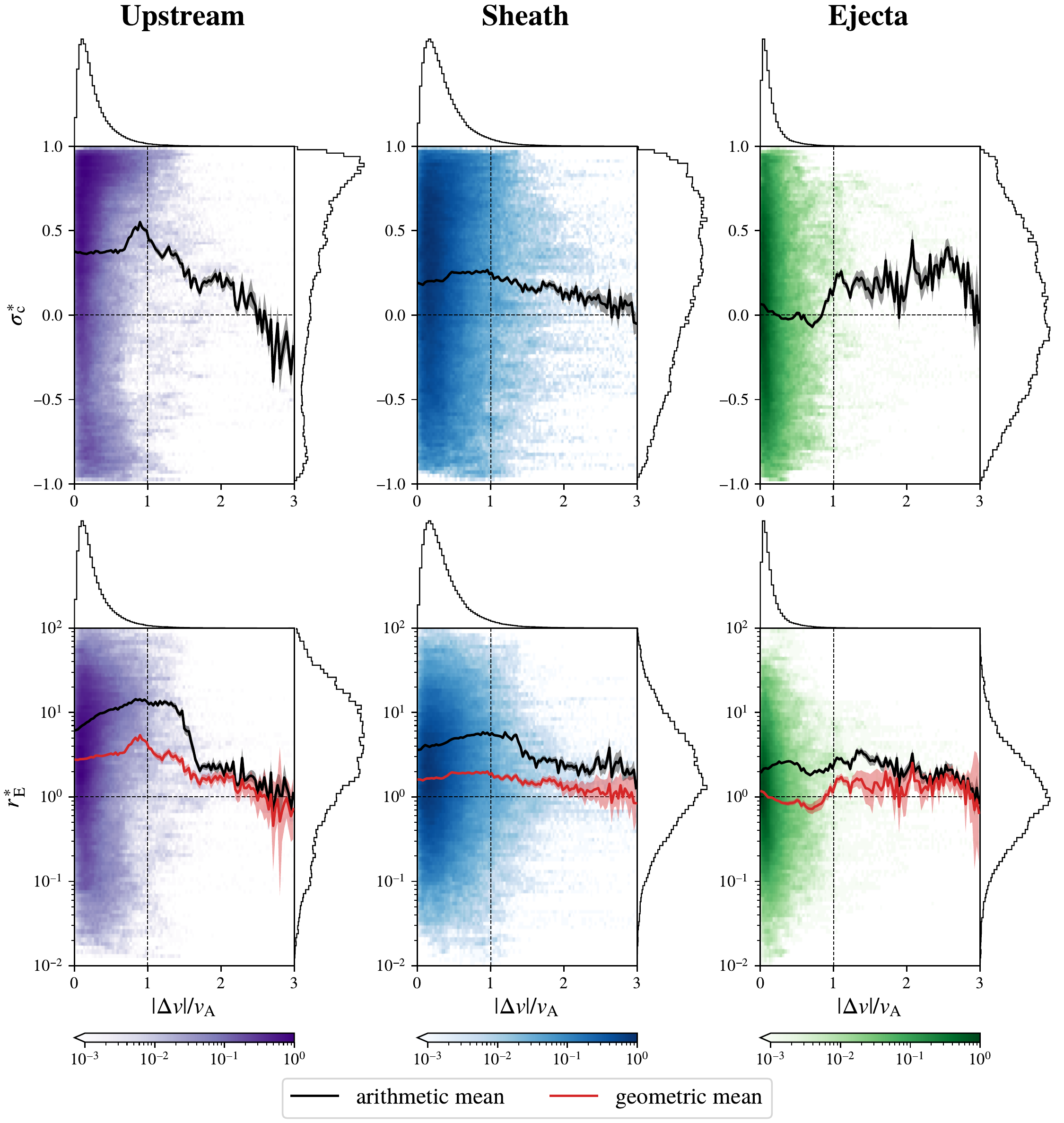}
    \caption{%
        2D distributions of \(\crosshelrect\) (top row) and \(\rErect\) (bottom
            row) against \(\KHI\) in the upstream solar wind, the sheath, and
            the ejecta, binned across all 74 events, with adjoining 1D
            histograms of \(\KHI\) (top panels), and \(\crosshelrect\) and
            \(\rErect\) (right-hand panels).
        The 2D histograms have been normalized to the maximum bin count.
        The black lines indicate mean \(\crosshelrect\) or \(\rErect\) within
            the vertical \(\KHI\) bins (calculated as an arithmetic mean).
        In the bottom row panels the red lines indicate mean \(\rErect\), but
            calculated as a geometric mean.
        The shaded error regions have been calculated as a standard error of the
            mean.}
    \label{fig:2d_distributions}
\end{figure*}

Overall the example event shows that individual sheaths have a lot of
    variability, which necessitates the use of statistical analysis to find
    general trends in the behavior of the different parameters.

\subsection{Parameter Distributions}\label{ssec:parameter_distributions}

Figure~\ref{fig:1d_distributions} shows the distributions of the residual
    energy, \(\resenerg\), the Alfvén ratio, \(\rA\), rectified cross helicity,
    \(\crosshelrect\), rectified Elsasser ratio, \(\rErect\), and the PSDs,
    \(\Epm\), of the Elsasser variables upstream of the shock, inside the
    sheath, and in the ejecta, summed across all 74 sheaths.
Both the upstream and ejecta intervals are \SI{8}{\hour} in duration.
In calculating these distributions the corresponding wavelet spectra were
    averaged across frequency, such that \(\resenerg\), \(\rA\),
    \(\crosshelrect\), and \(\rErect\) were calculated first and then averaged
    over frequency, rather than using the averaged PSDs to calculate the
    turbulence parameters.

In the top panels of Figure~\ref{fig:1d_distributions} the \(\resenerg\) and
    \(\rA\) distributions slightly shift to lower values, i.e., greater
    dominance of energy in \(\bm{b}\) fluctuations, when moving from the
    upstream wind into the sheath and from there to the ICME ejecta.
The average values of \(\resenerg\) of the three distributions are \num{-0.38},
    \num{-0.41}, and \num{-0.47}, respectively.
The upstream and sheath values line up well with earlier findings by
    \citet{chen2013} and \citet{good2022}, but the ejecta value is lower than
    the one found by \citet{good2022}, possibly due to the different choice of
    \(\bm{b}\) normalization.

\begin{figure*}
    \centering
    \includegraphics[width=\textwidth]{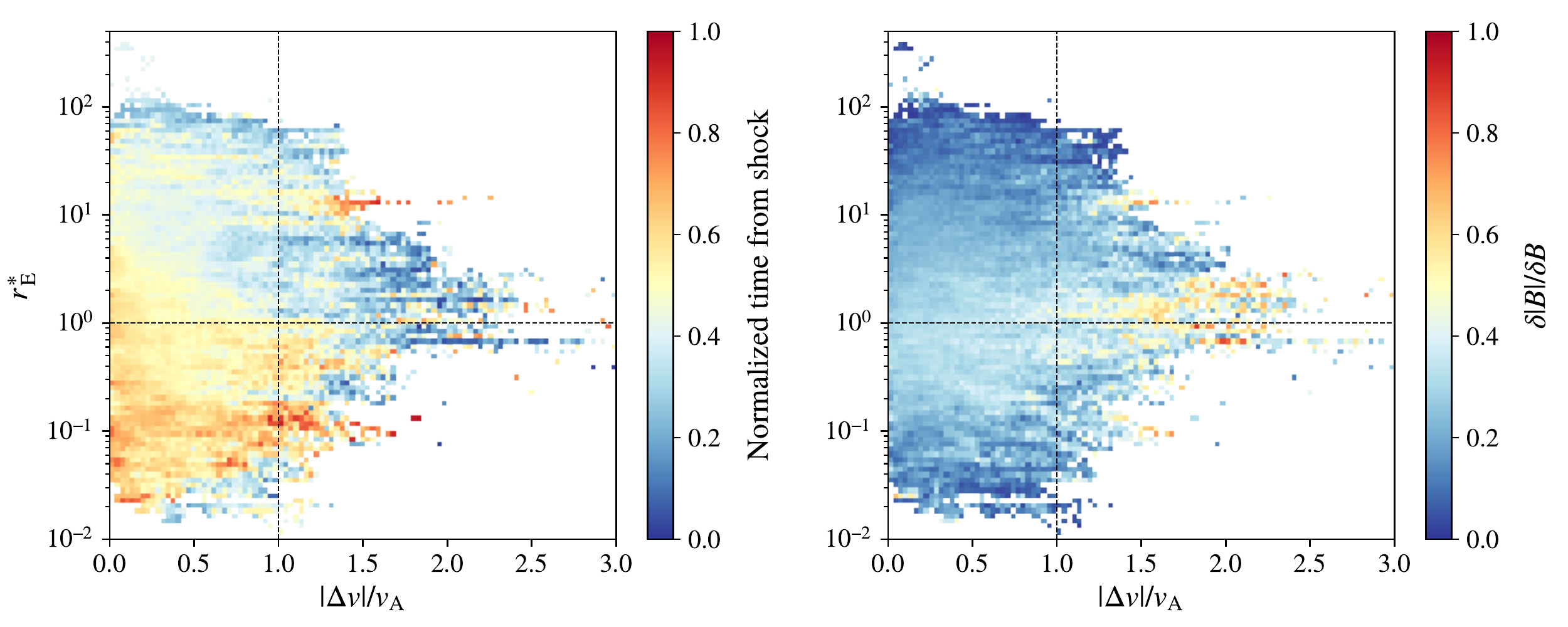}
    \caption{%
        2D distributions of \(\rErect\) against \(\KHI\), color-coded by the
            mean normalized time from shock (left) and \(\Bcomp\) (right) of
            values in each bin.
        In both histograms, bins containing fewer than five points have been
            excluded.}
    \label{fig:weighted_distributions}
\end{figure*}

The middle left panel of Figure~\ref{fig:1d_distributions} shows that
    \(\crosshelrect\) is mostly positive in the upstream solar wind, with a
    sharp peak near \(+1\).
This corresponds to the undisturbed solar wind at \SI{1}{\astronomicalunit}
    containing Alfvénic fluctuations propagating dominantly antisunward, which
    is consistent with earlier findings \citep[e.g.,][]{roberts1987a,chen2013}.
The distributions for the sheath and ejecta intervals are flatter,
    with considerably more negative \(\crosshelrect\) values than in
    the upstream distribution.
The ejecta distribution is symmetric and has a peak around \(\crosshelrect
    \approx 0\), indicating an overall tendency for a balance of sunward and
    antisunward fluctuations in the ICME ejecta.
This is consistent with~\citet{good2020b}.
The sheath distribution, while more balanced than the upstream, is tipped
    toward positive \(\crosshelrect\) values.
In the middle right panel, the \(\rErect\) distributions mirror the
    \(\crosshelrect\) behavior.
The upstream distribution is clearly weighted to the right, with a majority of 
    points having \(\rErect > 1\), indicating a dominance of antisunward
    fluctuations, while the ejecta distribution peaks at \(\rErect \approx
    1\), indicating sunward--antisunward balance.

In the bottom left panel, the three distributions for \(\Eplusrect\) are
    relatively similar, with the \(\Eplusrect\) sheath distribution having the
    highest mean value.
Greater differences are seen in \(\Eminusrect\), with the upstream wind
    distribution having a significantly lower mean value than the sheath and
    ejecta distributions.
While both \(\Eplusrect\) and \(\Eminusrect\) increase from the upstream wind
    to the sheath, it can be seen that the more balanced values of
    \(\crosshelrect\) and \(\rErect\) in the sheaths are caused by a relatively
    greater increase in \(\Eminusrect\).\vspace{3ex}

\subsection{Velocity Shear Dependence}\label{ssec:velocity_shear_dependence}

To investigate the possible link between velocity shear and the generation of
    sunward fluctuations, 2D histograms of \(\crosshelrect\) and \(\rErect\)
    versus the KH instability parameter, \(\KHI\), were
    calculated for the upstream solar wind, the sheath, and the ICME ejecta,
    with binning across all 74 events studied.
Figure~\ref{fig:2d_distributions} shows these distributions, along with
    1D histograms of the variables on the top and right-hand side adjoining
    axes.
The \(\KHI\) distributions were calculated using a bin width of \num{0.03}.
In addition to the 2D histograms, average values across the \(\crosshelrect\)
    and \(\rErect\) bins spanning each \(\KHI\) bin were calculated.
These averages are shown as black and red lines on top of the 2D histograms in 
    Figure~\ref{fig:2d_distributions}, with black and red 
    lines being arithmetic and geometric means, respectively.
The geometric mean of \(\rErect\) is mathematically similar to the arithmetic
    mean of \(\crosshelrect\).

While containing fewer points than the \(\KHI < 1\) parts of the distributions,
    there are sufficient data points at \(\KHI > 1\) (\num{14551} in the
    upstream, \num{37944} in the sheath, and \num{4899} in the ejecta) for
    robust statistical trends to be determined.
It is notable that velocity shears exceeding the linear KH instability occur
    relatively more frequently in sheaths than in the solar wind or ICME ejecta.
In the upstream wind, \(\crosshelrect \sim 0.4\) when \(\KHI < 1\), with a small
    increase in \(\crosshelrect\) seen just below the KH
    threshold, while the arithmetic mean of \(\rErect\) increases more smoothly
    across the interval (cf. the weaker rise in the geometric mean).
The behavior of the average \(\crosshelrect\) and \(\rErect\) lines changes when
    moving from the \(\KHI < 1\) region of the distribution to the \(\KHI > 1\)
    region.
At \(\KHI > 1\), \(\crosshelrect\) and \(\rErect\) show a decreasing trend with
    increasing \(\KHI\), with \(\crosshelrect\) and \(\rErect\) tending toward
    greater balance, i.e., values of zero and one, respectively.
The general behavior of the average \(\crosshelrect\) and \(\rErect\) inside the
    sheaths follows the same trend as in the upstream wind, i.e., invariance or
    relatively weak increase as \(\KHI\) grows in the \(\KHI < 1\) part of the
    distribution, and decrease toward balance when \(\KHI > 1\).

In the ICME ejecta, the average values of \(\crosshelrect\) and \(\rErect\)
    stay approximately constant and near balance when \(\KHI < 1\).
When \(\KHI > 1\) in the ejecta, uncertainties become more significant, such
    that \(\crosshelrect\) is positive but highly variable, and the geometric
    mean of \(\rErect\) is close to one but also varies significantly.

We also investigated the relationship between \(\rErect\) and \(\KHI\) inside
    the sheaths in terms of location within the sheath, as well as the link
    between shear and \(\Bcomp\).
The sheaths have been chosen for this analysis because, of the three interval
    types, they contained a greater abundance of large-amplitude shears.
In Figure~\ref{fig:weighted_distributions}, the bottom middle distribution from
    Figure~\ref{fig:2d_distributions} is reproduced, but here with color-coding
    by the average time from the shock (left panel) and average \(\Bcomp\)
    (right panel) of points in each bin; since the sheaths ranged in duration
    from \(\sim 3\) to \SI{22}{\hour}, the sheath durations were normalized
    such that zero refers to the shock time and one to the ICME leading edge
    time in the left panel.
    
At \(\KHI < 1\) in the left panel, there appears to be some correlation between
    \(\rErect\) and location in the sheath: fluctuations with the largest
    antisunward (sunward) imbalance tend to be located at the front (back) of
    the sheath, with the most balanced fluctuations tending to be found near
    the sheath midpoint.
This correlation is also present at \(\KHI > 1\) to a somewhat lesser degree.
Some of the largest shears (\(\KHI \gtrsim 1.75\)) are located near the front
    of the sheath, and are possibly generated by strong amplification of the
    nonradial velocity components by the shock (e.g., as seen in the
    Figure~\ref{fig:example_icme} example).

In the right panel the \(\rErect > \num{e1}\) and \(\rErect < \num{e-1}\)
    parts of the distribution are associated with very low \(\Bcomp\), with
    \(\Bcomp \lesssim 0.2\), while the middle part of the distribution is
    associated with somewhat higher \(\Bcomp\).
The highest values of \(\Bcomp\) are seen in the \(\KHI > 1\) and \(\rErect
    \sim 1\) part of the distribution, but there are no significant regions with
    \(\Bcomp \gtrsim 0.5\).

\begin{figure*}
    \centering
    \includegraphics[width=\textwidth]{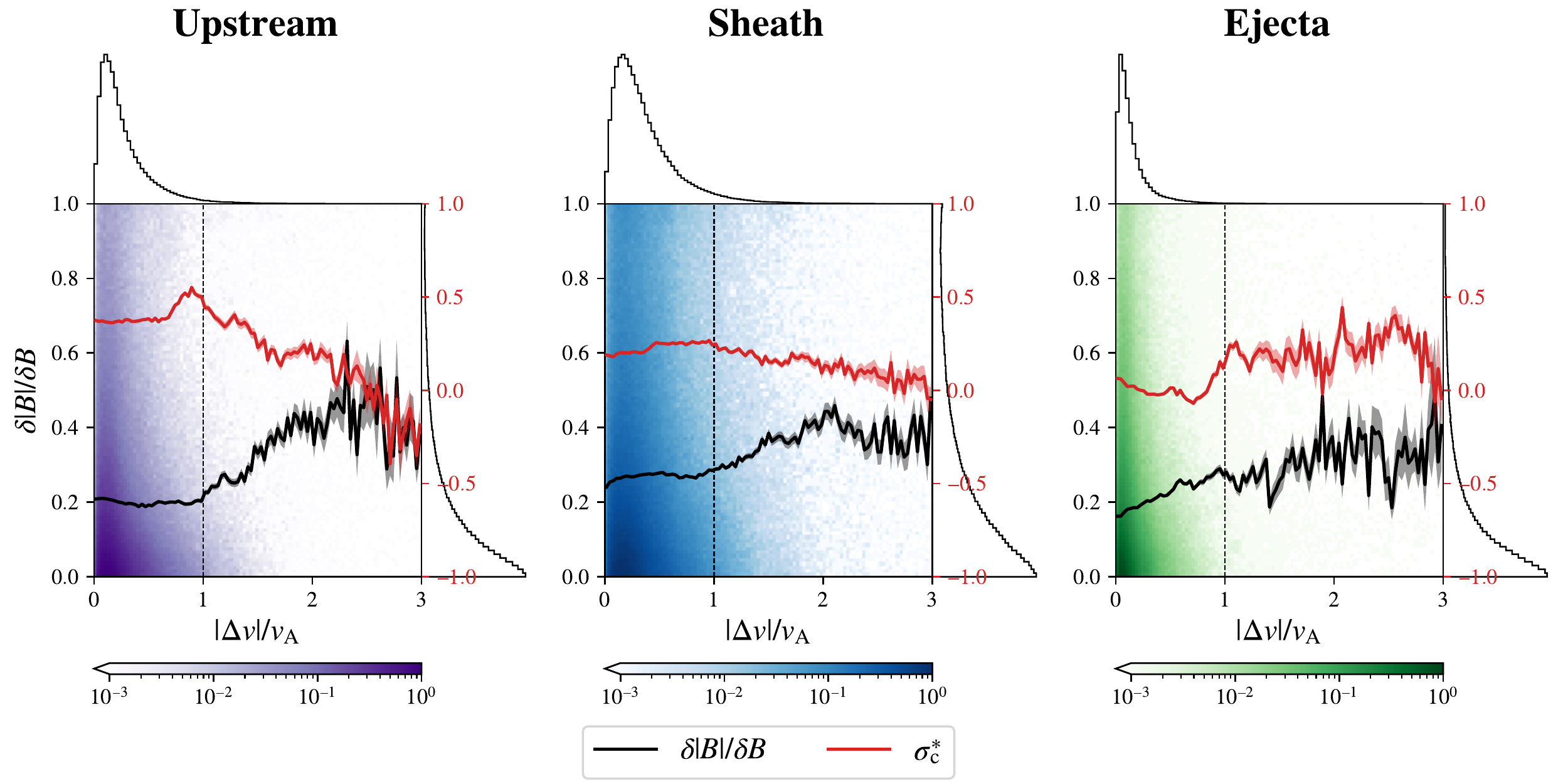}
    \caption{%
        2D distributions of \(\Bcomp\) against \(\KHI\) in the upstream
            solar wind, the sheath, and the ICME ejecta, binned across all 74
            events.
        Similar to Figure~\ref{fig:2d_distributions}, corresponding 1D
            distributions are included, the 2D histograms have been normalized
            to the maximum bin count, and the black lines indicate mean
            \(\Bcomp\) with corresponding error estimates.
        For comparison, the mean \(\crosshelrect\) line from
            Figure~\ref{fig:2d_distributions} has been superimposed over the
            distributions in red.}
    \label{fig:2d_comp_B_distributions}
\end{figure*}

To investigate how \(\Bcomp\) behaves with increasing \(\KHI\), 2D histograms
    of \(\Bcomp\) against \(\KHI\), similar to the ones in 
    Figure~\ref{fig:2d_distributions}, were calculated.
These are shown in Figure~\ref{fig:2d_comp_B_distributions}, with accompanying
    mean lines for \(\Bcomp\) as well as mean lines for \(\crosshelrect\) added
    on top to compare the behavior of the two parameters. 
In both the upstream solar wind and the sheath, \(\Bcomp\) remains fairly
    constant in the \(\KHI < 1\) part of the distribution, having a mean value
    of \num{0.20} in the upstream wind and \num{0.27} in the sheath.
At around the \(\KHI = 1\) threshold, \(\Bcomp\) starts increasing in both
    distributions.
Above \(\KHI \sim 2\), \(\Bcomp\) plateaus at around \num{0.3} to \num{0.4},
    with increased uncertainty in the values.

%%%%%%%%%%%%%%%%%%%%%%%%%%%%%%%%%%%%%%%%%%%%%%%%%%%%%%%%%%%%%%%%%%%%%%%%%%%%%%%%
\section{Discussion and Conclusion}\label{sec:discussion_conclusion}

In this Letter, we have presented a statistical analysis of the relation
    between turbulent imbalance and large-scale velocity shear in 74 ICME
    sheaths and their surroundings as observed by the \emph{Wind} spacecraft at
    \SI{1}{\astronomicalunit}.
When compared to upstream solar wind and downstream ejecta intervals, we found
    that sheaths contain significantly more large-amplitude shears exceeding
    the KH instability threshold, making sheaths particularly useful for
    studying the imbalance vs.\ shear relationship.

Firstly, there are clear differences in the overall distributions of the
    Elsasser ratio, \(\rErect\), and cross helicity, \(\crosshelrect\), between
    the three types of solar wind analyzed (Figure~\ref{fig:1d_distributions}).
Moving from the upstream wind into the sheath and then into the ICME ejecta,
    the mean of the \(\rErect\) distribution approaches one, and the mean of
    the \(\crosshelrect\) distribution approaches zero: these trends indicate a
    tendency toward balance between the sunward and antisunward Alfvénic
    fluctuations present in the plasma.

The observed shift toward higher \(\Epmrect\) combined with more balanced
    \(\rErect\) and \(\crosshelrect\) could be due to two mechanisms.
The first possibility is that, when swept into sheaths from the upstream wind,
    preexisting fluctuations are amplified without any significant generation
    of new fluctuations inside the sheaths.
However, this would require significantly stronger amplification of sunward
    than antisunward fluctuations to produce a reduction of \(\rErect\).
Alternatively, and perhaps more likely, new fluctuations are generated inside
    sheaths, with equal generation of \(\Epmrect\) that over time acts to shift
    the overall \(\rErect\) and \(\crosshelrect\) distributions toward balance
    \citep{smith2011} and away from the imbalance of the amplified, preexisting
    fluctuations that originated in the upstream wind.

In the upstream wind and sheaths, there is a clear association between velocity
    shear amplitudes at a 30-min timescale that exceed the KH instability
    threshold and more balanced fluctuations.
The fact that the behavior of the mean \(\crosshelrect\) and \(\rErect\) in
    Figure~\ref{fig:2d_distributions} is different above and below the \(\KHI =
    1\) threshold (relative invariance below, tending toward balance with
    increasing shear amplitude above) suggests that the KH instability can
    directly affect the balance of the turbulence.
The velocity shear vs.\ imbalance relationship is less clear inside the ejecta,
    possibly due to a combination of the ejecta plasma being intrinsically more
    balanced \citep{good2020b,good2022} and a relative scarcity of \(\KHI > 1\)
    shears.

We speculate that the shift toward balance at \(\KHI < 1\) in the distributions
    moving from upstream to sheath could be due to the prior action of velocity
    shears that are no longer present at \SI{1}{\astronomicalunit}, to some
    spreading of locally generated fluctuations outside of the \(\KHI > 1\)
    regions, or to the whole of the sheath intervals acting as velocity shears
    with \(\tau_{\mathrm{s}}\) equal to the sheath duration.
Near-Sun observations by \emph{Parker Solar Probe} and \emph{Solar Orbiter}
    could shed light on the first of these possibilities.

It should be expected that the dynamics of the velocity shear interface are
    nonlinear in nature, but the \(\KHI = 1\) threshold of the magnetized KH
    instability is a linear criterion.
This is still valid as an identifier of the onset of KH instability, as the
    early part of the dynamics can resemble a linear instability, which then
    evolves into a nonlinear instability with the associated vortex rollup
    \citep{ruffolo2020}.

There is also an association between velocity shear and magnetic
    compressibility that is clearly present in the upstream wind and sheath
    distributions.
Similar to the bimodal behavior of \(\crosshelrect\) and \(\rErect\), \(\Bcomp\)
    is flat below \(\KHI = 1\) and increases with shear amplitude above it.
This suggests that the KH instability also plays a role in generating
    compressive as well as Alfvénic fluctuations.

%%%%%%%%%%%%%%%%%%%%%%%%%%%%%%%%%%%%%%%%%%%%%%%%%%%%%%%%%%%%%%%%%%%%%%%%%%%%%%%%
\begin{acknowledgments}
    We thank the \textit{Wind} instrument teams for the data used in this study.
    We also thank the anonymous reviewer for their constructive comments on the
        manuscript.
    This work has been supported by the European Research Council under the
        European Union’s Horizon 2020 research and innovation programme, grant
        724391 (SolMAG), and by the B. E. Seljo fund of the Finnish Cultural
        Foundation, grant 00220897.
    SWG is supported by Academy of Finland Fellowship grants 338486 and 346612
        (INERTUM).
    The authors acknowledge additional support from Academy of Finland Centre of
        Excellence FORESAIL, grant 336807.
    We also wish to thank Vertti Tarvus for useful discussions.
    
    The Python wavelet analysis code, which is provided by Evgeniya Predybaylo
        and based on the work of \citet{torrence1998}, is available at
        \url{https://github.com/chris-torrence/wavelets}.
    The code to produce Figures~\ref{fig:2d_distributions}
        and~\ref{fig:2d_comp_B_distributions} was adapted from code originally
        developed by~\citet{stansby2019}.
\end{acknowledgments}

%%%%%%%%%%%%%%%%%%%%%%%%%%%%%%%%%%%%%%%%%%%%%%%%%%%%%%%%%%%%%%%%%%%%%%%%%%%%%%%%
\bibliography{references}{}


\end{document}